\documentclass[12pt]{article}
\usepackage{amsmath}
\usepackage{amscd}
\usepackage{array}
\usepackage{latexsym,amssymb}

\def\to{\rightarrow}

\newcommand{\eq}[1]{(\ref{#1})}

\newcommand{\be}{\begin{equation}}
\newcommand{\ee}{\end{equation}}
\newcommand{\bea}{\begin{eqnarray}}
\newcommand{\eea}{\end{eqnarray}}

\newcommand{\ba}{\begin{eqnarray}}
\newcommand{\ea}{\end{eqnarray}}

\def\ii{\mathrm{i}}

\def\notin{\hbox{{$\in$}\kern-.51em\hbox{/}}}

\def\inbar{\vrule height1.5ex width.4pt depth0pt}
\def\IB{\relax{\rm I\kern-.18em B}}
\def\IC{\relax\,\hbox{$\inbar\kern-.3em{\rm C}$}}
\def\ID{\relax{\rm I\kern-.18em D}}
\def\IE{\relax{\rm I\kern-.18em E}}
\def\IF{\relax{\rm I\kern-.18em F}}
\def\IG{\relax\,\hbox{$\inbar\kern-.3em{\rm G}$}}
\def\IH{\relax{\rm I\kern-.18em H}}
\def\II{\relax{\rm I\kern-.17em I}}
\def\IK{\relax{\rm I\kern-.18em K}}
\def\IL{\relax{\rm I\kern-.18em L}}
\def\IN{\relax{\rm I\kern-.18em N}}
\def\IP{\relax{\rm I\kern-.18em P}}
\def\IQ{\relax\,\hbox{$\inbar\kern-.3em{\rm Q}$}}
\def\IR{\relax{\rm I\kern-.18em R}}
\def\IU{\relax\,\hbox{$\inbar\kern-.3em{\rm U}$}}
\def\ZZ{\relax\ifmmode\mathchoice{\hbox{\cmss
Z\kern-.4em Z}}{\hbox{\cmss Z\kern-.4em Z}}{\lower.9pt\hbox{\cmsss Z\kern-.4em Z}} {\lower1.2pt\hbox{\cmsss
Z\kern-.4em Z}}\else{\cmss Z\kern-.4em Z}\fi}
\def\IGam{\relax{{\rm I}\kern-.18em \Gamma}}

\def\bfnull{\relax{\rm O \kern-.635em 0}}

\def\square{{\,\lower0.9pt\vbox{\hrule
\hbox{\vrule height 0.2 cm \hskip 0.2 cm \vrule height 0.2 cm}\hrule}\,}}

\def\twomat#1#2#3#4{\left(\begin{array}{cc} \end{array} \right)}

\begin{document}

\numberwithin{equation}{section}
\begin{flushright}
CERN Preprint 2012-002
\end{flushright}

\vskip 1.5cm

\begin{center}
{\Large Black holes in the Superworld } \\

\vskip 1.5cm

{\bf \large Laura Andrianopoli$^{1}$, Riccardo D'Auria$^1$\\
  and Sergio Ferrara$^{2}$}\\
\vskip 8mm
 \end{center}
\noindent {\small{\it $^1$  Dipartimento di Fisica, Politecnico di Torino, Corso Duca
    degli Abruzzi 24, I-10129 Turin, Italy and Istituto Nazionale di
    Fisica Nucleare (INFN) Sezione di Torino, Italy; E-mail:  {\tt
      laura.andrianopoli@polito.it}; {\tt riccardo.dauria@polito.it}; {\tt mario.trigiante@polito.it}\\
    $^2$
    Physics Department, theory Unit, CERN, 1211 Geneva 23, Switzerland and INFN LNF,
    Via E. Fermi 40 00044 Frascati, Italy;
     E-mail: {\tt sergio.ferrara@cern.ch}}}
\vfill
\begin{center}
{\bf Abstract}
\end{center}
Some aspects  of black holes in supersymmetric theories of gravity are reviewed and some recent results
outlined.

\vfill {\small Contribution to the Proceedings  of the International Symposium on ``Subnuclear Physics: Past,
Present and Future" held at the Pontificial Academy of Sciences, Vatican City 30 October - 2 November 2011,
based on a talk given by Sergio Ferrara.}

\eject




\section{Introduction}
\label{intro} Black holes are perhaps the most misterious and fascinating outcome of Einstein's theory of
General Relativity (\emph{A.~Einstein, 1880-1952}). This theory was the result of a deep intuition on the
implications of the equivalence principle,
 whilst trying to merge Newton's Law of gravitation with general relativistic covariance.
 Its mathematical formulation was then realized in terms of Riemannian geometry of space-time (\emph{B.~Riemann, 1826-1866}).
 Nowadays
black holes are predicted by fundamental candidate theories of Quantum Gravity like Superstring or M-Theory and
they are observed in the sky as relics of collapsing stars. They seem to encompass many of the mysteries of the
evolution of our Universe from its creation to its final destiny, the so-called Big Crunch,  or its eternal
existence, namely an endless expansion.

Astrophysical black holes have huge masses, typically of the order of magnitude of the solar mass scale,
$2\times 10^{30}$ Kg, while Quantum Gravity black holes have tiny masses, of the order of the Planck mass scale,
namely $2\times 10^{-8}$ Kg. note that this is still much bigger than the typical mass of the atomic nuclei,
that is from one to ten proton masses (the mass of a proton being $1.6\times 10^{-27} $ Kg).

Supergravity black holes are the black holes of the superworld \cite{ADFT}. Supersymmetry requires that they are
extremal, that is that they have vanishing temperature, are marginally stable but carry Entropy. Actually, the
black-hole Entropy makes a bridge between classical gravity and Quantum Gravity. In fact, we recall that the
macroscopic definition of the black-hole entropy (Bekenstein -- Hawking Entropy) \cite{haw,bek} connects its
value to the black-hole horizon area $A_H$:
\begin{equation}
S^{macro}_{BH}= \frac{k_B c^3}{G \hbar}\frac 14 A_H
\label{SBH}
\end{equation}
The microscopic definition of the black-hole entropy, instead, relates its value to the number $N_{\mbox{\tiny
mic}}$ of microstates of the quantum system underlying the black hole, namely:
\begin{equation}
S^{micro}_{BH}=  k_B \ln (N_{\mbox{\tiny mic}})
\end{equation}
Remarkably these formulae, computed with appropriate approximations, give the same result in Superstring Theory
\cite{SV}.

From now on, we generally use Natural Units, where $c=\hbar=G=k_B=1$.

\section{What is the Superworld?}
In order to understand what Superworld is, one first has to introduce the notion of Superspace \cite{SS,FWZ}.
This is a geometrical entity which extends the notion of Riemannian manifold to that of Supermanifold. A point
on a $D$-dimensional Riemannian manifold $\mathcal{M}_D$, endowed with a Lorentz signature (\emph{H.~A.~Lorentz,
1853-1928}), is identified by giving numerical coordinates $x^\mu$, ($\mu = 1,\cdots , D$). To identify a point
in a Supermanifold we need, besides the coordinates $x^\mu$, also a set of Grassmann (\emph{H.~Grassmann,
1809-1877}) anticommuting coordinates
 $\theta_\alpha$ ($\alpha =1,\cdots , 2^{[D/2]}$) with two basic properties:
\begin{enumerate}
\item
$\theta_\alpha \theta_\beta = -\theta_\beta \theta_\alpha$\\ which implies nilpotency: $\theta_\alpha^2=0$;
\item
They transform as spinors (\emph{E.~Cartan, 1869-1951}, \emph{H.~Weyl, 1885-1955}) under the action of the
Lorentz group and their properties are related to modules of Clifford  Algebras (\emph{W.~K.~Clifford,
1845-1879}) and to the Spin Group, namely the universal covering group of the $D$-dimensional Lorentz group
 \cite{del}.
\end{enumerate}

 Superworld is the  physical entity
 corresponding to the mathematical concept of supermanifold, whose environment is not ordinary space but superspace.
The group of motion in Superspace is supersymmetry, as much as the group of motion in ordinary space-time is the
Poincar\'e group (\emph{H.~Poincar\'e, 1854-1912}). An infinitesimal supersymmetry transformation with spinorial
parameter $\epsilon_\alpha$ acts on the coordinates of superspace as follows:
\begin{eqnarray}
x^\mu \to x^\mu + \ii \bar\epsilon^\alpha (\gamma^\mu)_\alpha{}^\beta \theta_\beta  &\Leftrightarrow
& \delta x^\mu =\ii \bar\epsilon^\alpha (\gamma^\mu)_\alpha{}^\beta \theta_\beta \\
\theta_\alpha \to  \theta_\alpha + \epsilon_\alpha &\Leftrightarrow
& \delta\theta_\alpha =\epsilon_\alpha
\end{eqnarray}
where $\gamma^\mu$ is a matrix satisfying the Clifford Algebra and $\bar\epsilon$ denotes the Dirac conjugate
spinor, namely $\bar\epsilon= \epsilon^\dag \gamma^0$.
 Commuting twice the action of such
transformations with parameters $\epsilon_1$ and $\epsilon_2$ respectively, one finds that $x^\mu$ undergoes an
infinitesimal translation:
\begin{equation}
[\delta_1,\delta_2] x^\mu = 2 \ii \bar\epsilon_{2}^\alpha (\gamma^\mu)_\alpha{}^\beta \epsilon_{1\beta}\,.
\end{equation}
The supersymmetry algebra is a graded Lie algebra \cite{CNS} (\emph{S.~Lie, 1842-1899}) with basic
anticommutator: \cite{WZ,GL,VA}
\begin{equation}
\{Q_\alpha,\bar Q_\beta\}= 2 (\gamma^\mu C)_{\alpha\beta} p_\mu
\end{equation}
where the supersymmetry generators $Q_\alpha$ are Majorana spinors (\emph{E.~Majorana, 1906-1938}) and $C$
denotes the charge-conjugation matrix.

The supergroup associated to the supersymmetry algebra acts on a supermanifold, denoted by
$\mathcal{M}_{b,f}\equiv \mathcal{M}_{D,2^{[D/2]}} $, where $b$ and $f$ denote the number of bosonic and
fermionic coordinates respectively. the total (graded) dimension of a supermanifold is $b+f$. As we will see in
the following, the maximal possible number of coordinates of superspace is $b_{max}=11$, $f_{max}= 32$.

Actually, one can extend the super Lie algebra by introducing $N$ supersymmetry generators $Q_{\alpha I}$
($I=1,\cdots,N$) acting on an $N$-extended superspace labeled by $N$ Grassmannian spinor coordinates
$\theta_{\alpha I}$. The basic anticommutators now become
\begin{eqnarray}
\{Q_{\alpha I},\bar Q^J_\beta\}&=& 2 (\gamma^\mu C)_{\alpha\beta} p_\mu \delta_I^J\\
\{Q_{\alpha I}, Q_{\beta J}\}&=& \varepsilon_{\alpha\beta} Z_{IJ}
\end{eqnarray}
where $Z_{IJ}$ are ``central terms" which commute with all the rest of the superalgebra, including the Lorentz
generators. It is precisely the presence of the central charges $Z_{IJ}$ which makes it possible the existence
of supersymmetric Black Holes, as will be shown in the next section.

The interaction in the superworld are described by supersymmetric theories. It is remarkable that such theories
may encompass gauge interactions, in particular Yang--Mills theories \cite{FZ,SS}, as well as gravitational
interactions. In the latter case, the corresponding theory is called supergravity \cite{FFV,DZ}. However these
theories exist only for few values of the number $N$ of supersymmetries and of the space-time dimension $D$
\cite{G,N}. In particular, Super Yang--Mills theories in $D=4$ require $1\leq N\leq 4$ and at most they live in
$D=10$ \cite{BSS}. On the other hand supergravity theories at $D=4$ require $1\leq N\leq 8$ and at most they
live in $D=11$ dimensions \cite{CJS}.

\section{From Schwarzschild to Reissner--Nordstr\"om: The case of extremal Black Holes.}

The celebrated Black-Hole solution of pure Einstein theory looks, in a chosen frame of spherical coordinates
\begin{equation}
ds^2_{\mbox{\tiny Schw}}= -\left(1-\frac{2M}r\right) dt^2 +\left(1-\frac{2M}r\right)^{-1} dr^2 + r^2 d\Omega^2 \,
\end{equation}
where $M$ denotes the ADM mass, that is the total energy of the black-hole configuration. The naked singularity
at $r=0$ is covered by the \emph{event horizon} at $r=2M$. By event horizon we mean a surface where the
gravitational red-shift is infinite, that is where the time intervals undergo an infinite dilation with respect
to a distant observer. This is also a singularity of the metric but it is only a \emph{coordinate} singularity
which can be removed with an appropriate choice of coordinates, while the singularity at $r=0$ is a real
singularity of the theory, that is independent of the reference frame.

The generalization of the Schwarzschild solution to an electrically charged black hole in the Einstein--Maxwell
theory is given by the Reissner--Nordstr\"om black hole, whose metric reads:
\begin{equation}
ds^2_{\mbox{\tiny RN}}= -\left(1-\frac{2M}r+\frac{Q^2}{r^2}\right) dt^2 +\left(1-\frac{2M}r+\frac{Q^2}{r^2}\right)^{-1} dr^2 + r^2 d\Omega^2
\label{rn}
\end{equation}
Here $M$ denotes the ADM mass and $Q$ the electric charge of the space-time configuration. Such configuration
can be easily generalized for dyonic configurations where also a magnetic charge $P$ is present by replacing in
\eq{rn} $Q^2$ with $Q^2 + P^2$.
 This metric exhibits two horizons, at
\begin{equation}
 r_\pm = M \pm \sqrt{M^2-Q^2}= M\pm r_0
 \end{equation}
 where $r_+$ corresponds to the event horizon and $r_-$ to the Cauchy horizon, together with the
  physical singularity at $r=0$.

 In cosmology a Cosmic Censorship Principle is postulated (see for example \cite{KL}.
 An event horizon should always cover the singularity at $r=0$, so that the singularity be not accessible to an
 observer external to the event horizon of the black hole. This can be rephrased by saying that no "naked" singularities
  can exist. For the
Reissner--Nordstr\"om solution the Cosmic Censorship principle requires $M\geq Q$.

As shown by Steven Hawking,  black holes obey laws which are formally the same as the laws of thermodynamics,
after an appropriate identification of the quantum numbers of the solution is given. In particular, the
thermodynamical properties of the black holes relate the area of the event horizon to the Entropy  through the
Bekenstein--Hawking formula (see eq. (\ref{SBH}) in Natural Units):
\begin{equation}
S_{BH}= \frac 14 A_H =\frac 14 \pi R_+^2\,,
\end{equation}
where $R_+$ is the event horizon $r_+$ for the Reissner--Nordstr\"om black hole, while it becomes an effective
radius in the presence of other black-hole attributes such as angular momentum $J$ and/or scalar charges
$\Sigma$. For instance, in the presence of the latter $R_+^2 = r_+^2 -\Sigma^2\leq r_+^2$.

The fact that a black hole continuously increases its horizon area can be interpreted as the second law of
thermodynamics if we identify the black hole entropy as proportional to the horizon area, as pointed out by S.
Hawking. Further support to this interpretation is given by the other laws of thermodynamics. in particular, the
$0^{th}$ law of thermodynamics states that for a system in equilibrium there is a quantity, the temperature,
which is constant. An analogous constant quantity exists for a black hole at equilibrium, the so-called surface
gravity that for the Reissner--Nordstr\"om black hole is
\begin{equation}
\kappa= \frac{c}{(r_++r_-) r_+ - Q^2}
\end{equation}
where
\begin{equation}
c = \frac 12(r_+-r_-)
\end{equation}
It is then possible to identify the black-hole temperature $T_{BH}$ as
\begin{equation}
T_{BH}= \frac 1{2\pi} \kappa = \frac{c}{2 S_{BH}}
\end{equation}
The analogy is completed by rewriting the first law of thermodynamics:
\begin{equation} dE= TdS + \mbox{work
terms}
\end{equation}
as
\begin{equation}
 dM= T_{BH} dS_{BH} + \cdots = \frac{\kappa}{2\pi} \frac{A_H}{4} +\cdots
\end{equation}
and observing that the third law of thermodynamics, stating that it is impossible to achieve $T=0$ by a finite
number of physical processes, can be rephrased as the impossibility to achieve $\kappa =0$ by a finite number of
physical processes.

The black hole which reaches the limit equilibrium temperature $\kappa= 0$ is called extremal. This corresponds
to $c=0$, that is to $r_+=r_-$. For the Reissner--Nordstr\"om black hole, this happens when $M=|Q|$. A
supergravity black hole is supersymmetric (BPS saturated) if its ADM mass $M$ equals the highest skew-eigenvalue
of the central charge matrix $Z_{IJ}=-Z_{JI}$ evaluated at asymptotic infinity. In the presence of scalar
charges $\Sigma$, the extremality condition allows both for supersymmetric and non-supersymmetric black-hole
configurations.

For stationary but non-static solutions, that is rotating black holes of angular momentum $J$ (Kerr--Newman
black holes), the horizon radii become
\begin{equation}
r_\pm= M \pm \sqrt{M^2 -Q^2 -P^2 -J^2/M^2}
\end{equation}
so that for a neutral spinning black hole (Kerr black hole) we reach extremality when $M^2=J$, that is when the
extremality parameter $a^*\equiv J/(GM^2)=1$. Kerr black holes have been observed in our galaxy, in particular
\textbf{GRS 1915+105} is the heaviest of the stellar black holes so far known \cite{McC} in the Milky Way
Galaxy, with 10 to 18 times the mass of the Sun and a value of spin  $J\simeq 10^{78} \hbar$. It was discovered
on 15 August 1992. It is a nearly extremal black hole since in this case the extremality parameter is $a^*= 0.98
\simeq 1$. It has been argued that such black hole has an exact Conformal Field Theory dual \cite{GHSS}.

\section{Black Holes and Supersymmetry}
One of the main properties of supergravity is the presence of scalar fields not minimally coupled to vector
fields. The typical form of the Lagrangian of a set of electromagnetic field strengths $F^\Lambda = d A^\Lambda$
(enumerated by capital Greek indices $\Lambda, \Sigma $) in supergravity is of the form:
\begin{equation}
\mathcal{L} \propto g_{\Lambda\Sigma}(\varphi) F^\Lambda_{\mu\nu} F^{\Sigma |\mu\nu} + \Theta_{\Lambda\Sigma} (\varphi)
\frac 12 F^\Lambda_{\mu\nu} F^{\Sigma}_{\rho \sigma} \epsilon_{\mu\nu\rho\sigma}
\end{equation}
where the couplings  $g_{\Lambda\Sigma}$ and $\Theta_{\Lambda\Sigma}$ depend on a set of scalar fields
enumerated by an index $s$. This has the implication that the Maxwell--Einstein black hole solution gets a
non-trivial modification. In particular, the metric flow of the black hole towards the horizon $r=r_H$ is
accompanied by trajectories of scalar-fields evolution from asymptotic infinity to the horizon:
\begin{eqnarray}
\lim_{r\to \infty}\varphi^s (r) &=& \varphi^s_\infty \in \mathcal{M}\nonumber\\
\lim_{r\to r_H}\varphi^s (r) &=& \varphi^s_{crit} \in \mathcal{M}
\end{eqnarray}
The resulting analysis exploits the \emph{attractor mechanism} \cite{FKS}. Indeed scalar fields behave as
dynamical systems which, in their evolution towards the black-hole horizon of an extremal black hole, loose
memory of their initial conditions (at $\varphi =\varphi_\infty$) approaching a critical point where the first
derivative vanishes:
\begin{eqnarray}
\lim_{r\to r_H}\varphi^s (r) &=& \varphi^s_{crit} (Q)\nonumber\\
\lim_{r\to \infty}\frac{d}{d r}\varphi^s (r) &=& 0\,,
\end{eqnarray}
and whose value only depends on the set of charges $Q$. Consistency of the solution implies that
$\varphi_{crit}$ is a critical point of an \emph{effective black-hole potential} $V_{BH}(\varphi, Q)$:
\begin{equation}
\lim_{r\to \infty}\frac{\partial}{\partial \varphi^s}V_{BH}|_{\varphi=\varphi_{crit}} = 0\,.
\end{equation}
Moreover, the Bekenstein--Hawking entropy-area formula becomes \cite{FK,CDF}:
\begin{equation}
S_{BH} = \frac 14 A_H = \pi V_{BH}(Q,\varphi_{crit}(Q))\,.
\end{equation}
Note that the critical value of the scalar fields for extremal black holes, satisfying the attractor mechanism,
is given only in terms of the electric and magnetic vector of charges $Q$, and this explains why the entropy
only depends on $Q$ and not on the initial values of the scalar fields $\varphi_{\infty}$.

The attractor mechanism allows to reduce the dynamical black-hole flow to a \emph{first-order} evolution both
for supersymmetric and non-supersymmetric extremal black holes \cite{CD,ADOT,Ceresole:2009iy,PB}. Indeed, the
black-hole potential $V_{BH}$ may be always written for extremal black holes as
\begin{equation}
V_{BH}= W^2 + 2 \partial_s W \partial^s W
\end{equation}
where $W(\varphi,Q)$ is known as the \emph{(fake) superpotential}. There are several properties of $W$ that make
it important. First of all, in terms of $W$, the second order equations of motion of the theory reduce to a set
of first order equations. Moreover $W$ is invariant under the electric-magnetic duality group. It has a clear
meaning in the context of the Hamilton--Jacobi theory, since it allows the interpretation of the flow as an
Hamiltonian flow whose Hamilton characteristic function is actually simply related to $W$ \cite{ADOT2,ADFT2}.

The attractor mechanism  allows to classify black-hole solutions, that is critical points of the black-hole
potential, through the electric-magnetic duality symmetry of the theory. For each orbit of the duality symmetry,
the fake superpotential $W$ has a different expression. In the particular case of supersymmetric black holes,
one obtains $W=|Z|$ where $|Z|$ is the highest skew-eigenvalue of the central charge matrix $Z_{IJ}$. The
duality orbits are modules of \emph{groups of type $E_7$}, as requested by the Gaillard--Zumino analysis
\cite{GZ} combined with the attractor mechanism. The group $E_7$ appears in supergravity as the duality group of
of the maximally extended $N=8$ theory in four dimensions, in its symplectic 56 dimensional module relating 28
electric to 28 magnetic charges. The orbits of the \textbf{56} module classify the black-hole solutions
preserving different fractions of the original $N=8$ supersymmetry. Moreover, $E_7$ controls the ultraviolet
divergences of perturbation theory since it is anomaly free, and its arithmetic subgroups $G\subset
E_7({\mathbb{Z}})$ may encode the non-perturbative quantum corrections. It happens that all the duality groups
of four dimensional supergravity theories with a number of supersymmetries $N<8$ are groups of type $E_7$, that
is they have symplectic representations admitting a symmetric quartic invariant polynomial, but not a quadratic
one \cite{FK2}.
\begin{table}[h]
\begin{center}
\begin{tabular}{|c|c|c|c|}
\hline
& $G$ &$R$ module & Primitive symm. inv.\\
\hline  \hline
$J^\mathbb{O}_3$ & $E_{7(-25)}$  & 56 & $I_4$\\
 \hline
$J^\mathbb{H}_3$ & $SO^*(12)$  & 32 & $I_4$\\
 \hline
$J^\mathbb{C}_3$ & $SU(3,3)$  & 20 & $I_4$\\
 \hline
$J^\mathbb{R}_3$ & $Sp(6,\mathbb{R})$  & $14'$ & $I_4$\\
 \hline
$T^3$ & $SL(2,\mathbb{R})$  & 4  & $I_4$\\
 \hline
$J_{2,n}$ & $SL(2,\mathbb{R}) \times SO(2,n)$  & $(2,2+n)$ & $I_4$\\
 \hline
$CP^n$ & $U(1,n)$  & $(1+n)_C$ & $I_2$\\
 \hline
\end{tabular}
\end{center}
\caption{Supergravity sequence for $N=2$ symmetric spaces.}
\end{table}
As an example, we have presented in Table 1 the possible  $N=2$ choices of groups $G$ of the ${G\over H}$
symmetric spaces and their symplectic representations $\mathbf{R}$ \cite{GST,CVP}. The first column identifies
the scalar manifold whose isometry group is $G$ is given. In particular, for the first four entries, they are
named with the  Jordan algebras $J_3$ over octonions, quaternions, complex and real numbers respectively, to
which they are related. As we see in the last column,  all the duality groups listed are of Type $E_7$ groups
\cite{BGM} except the last one, which has a primitive quadratic invariant polynomial. For  $N\geqslant 3$
supergravity theories the analogous sequence is given in Table 2. Again, all the groups appearing in the
sequence are of type $E_7$ except the first one, which admits a primitive symmetric invariant polynomial of
order 2.
\begin{table}[h]
\begin{center}
\begin{tabular}{|c||c|c|}
\hline
$N$& $
\begin{array}{c}
\\
$G$ \\
~
\end{array}
$ & $
\begin{array}{c}
\\
  \mathbf{ R}
\\
~
\end{array}

$ \\ \hline\hline $
\begin{array}{c}
\\
N=3 \\
~
\end{array}
$ & $U(3,n)$ & $ \mathbf{(3+n)}$     \\ \hline $
\begin{array}{c}
\\
N=4 \\
~
\end{array}
$ & $SL(2, \mathbb{R})\otimes {SO(6,n)}$ & $\mathbf{(2, 6+n)}$   \\ \hline $
\begin{array}{c}
\\
N=5 \\
~
\end{array}
$ & $SU(1,5)$ & $ \mathbf{ 20}$   \\ \hline $
\begin{array}{c}
\\
N=6 \\
~
\end{array}
$ & $SO^{\ast }(12)$ & $\mathbf{ 32}$
 \\ \hline
$
\begin{array}{c}
\\
N=8 \\
~
\end{array}
$ & $E_{7\left( 7\right) }$ & $\mathbf{ 56}$   \\ \hline
\end{tabular}
\end{center}
\caption{The supergravity sequence for $N\geq 3$}
\end{table}

\section{Future directions of research}
We are just at the beginning of the exploration of the beautiful intricacy given by supergravity black-hole
physics, its group-theoretical structure and quantum perspectives. It is clear that much work and effort has to
be done to unveil all the physics behind their structure which are emerging from supergravity considerations. In
particular, we may mention few possible future directions of research: \begin{itemize}
\item Extension of black-hole solutions to multi-center configurations, the classification of their orbits and
the study of their dynamics, regarding their splitting behavior and their relation to the underlying stringy
microstate counting.
\item Clarification of the role of the group $E_7$ as far as quantum corrections are concerned.
\item Inclusion of the Attractor Mechanism in the presence of higher derivative modifications of gravity, as
suggested by superstring theory.
\item The role of $N=8$ black holes in a perturbatively finite theory of $N=8$ supergravity.
\end{itemize}

\section*{Acknowledgements}
The work of S.F. is supported by the ERC Advanced Grant n. 226455, ``Supergravity, Quantum Gravity and Gauge
Fields". This work was supported in part by the MIUR-PRIN contract 2009-KHZKRX.

\end{document}